# On the Origin and Evolution of Life in the Galaxy

Michael McCabe and Holly Lucas

Department of Mathematics, University of Portsmouth,
Lion Terrace, Portsmouth, Hants PO41 3HF  UK

**Abstract**

A simple stochastic model for evolution, based upon the need to pass a sequence of *n* critical steps (Carter 1983, Watson 2008) is applied to both terrestrial and extraterrestrial origins of life. In the former case, the time at which humans have emerged during the habitable period of the Earth suggests a value of $n = 4$. Progressively adding earlier evolutionary transitions (Maynard Smith and Szathmary, 1995) gives an optimum fit when $n = 5$, implying either that their initial transitions are not critical or that habitability began around 6 Ga ago. The origin of life on Mars or elsewhere within the Solar System is excluded by the latter case and the simple anthropic argument is that extraterrestrial life is scarce in the Universe because it does not have time to evolve. Alternatively, the timescale can be extended if the migration of basic progenotic material to Earth is possible.  If extra transitions are included in the model to allow for Earth migration, then the start of habitability needs to be even earlier than 6 Ga ago. Our present understanding of Galactic habitability and dynamics does not exclude this possibility. We conclude that Galactic punctuated equilibrium (Cirkovic et al. 2009), proposed as a way round the anthropic problem, is not the only way of making life more common in the Galaxy.

**Keywords**

Evolution, origin of life, Galaxy, stochastic, mathematical model, critical evolutionary transitions, Galactic Habitable Zone, habitability, Earth migration

**Introduction**

It is usually accepted that our atomic origins are extraterrestrial and that physics is universal. The chemical elements required for life were not created on Earth, but through nucleosynthesis during the Big Bang (H,He,Li), inside the cores of stars (He, C, N, O …) and within supernovae (Ca, P, S, Na, K, Cl …). The origin of terrestrial molecules required for life, notably water, is less certain, but comets and meteorites impacting the young Earth must have made a significant contribution and extraterrestrial origins are certainly not excluded. Water ice and organic material has recently been detected on a main belt asteroid (Rivkin and Emery, 2010) increasing their possible role in delivering molecules required for life. Chemistry is regarded as universal and our molecular origins are accepted as being both terrestrial and extraterrestrial. Biology is not generally regarded as universal and the question of whether our biological origins are terrestrial or extraterrestrial is highly controversial (Table 1).

Standard evolutionary biology adopts a time on Earth between 3.5 and 4 Ga ago as its starting point with deep ocean hydrothermal vents as one possible location (Martin et al., 2008) and recognises that the rapid emergence of life is a puzzle. Maynard Smith and Szathmary (1995) have suggested eight steps on Earth leading from replicating molecules to mankind, of which seven are regarded as critical. Table 1 includes the critical transitions (1) to (7), but omits the faster transition expected from solitary individuals to colonies between (6) and (7).

| Entity | Modern Examples | ET Origin | Transition |
|---|---|---|---|
| small atoms | H to He | Big Bang | nucleosynthesis |
| medium atoms | Li to Fe | stellar cores | nucleosynthesis |
| large atoms | above Fe | supernovae | nucleosynthesis |
| smaller molecules | $H_2O$ | ISM, comets | chemical reaction |
| larger molecules | glycine | ISM, meteorites | chemical reaction |
| replicating molecules | lipids | comets/meteorites | chemical reaction |
| protocells | bilayer vesicles | least unlikely | self organisation (1) |
| chromosomes | retrovirus (RNA) | less unlikely | synthesis (2) |
| prokaryotes | bacteria | unlikely | RNA to DNA (3) |
| eukaryotes | algae | more unlikely | endosymbiosis (4) |
| protists | protozoa | most unlikely | asexual to sexual (5) |
| animals/plants/fungi | primates | science fiction | cell differentiation (6) |
| humans | authors MM/HL | science fiction | language (7) |

Table 1 Extraterrestrial Origins and Transitions Required for Human Life to Emerge

A simple stochastic model, first proposed by Carter (1983) and discussed by Barrow and Tipler (1986), Hanson (1998), Flambaum (2003), assumes that evolution is governed by a series or critical transitions or bottlenecks, which must be passed to advance from basic molecules to simple, complex and then intelligent life. The anthropic model allowed the derivation of a probability density function and expectation time for the final step, based upon the habitable lifetime of the planet. Watson (2008) has further analysed the model to derive probability density functions and expectation times for every step. He identifies seven critical transitions in biological evolution from amongst the eight proposed by Maynard Smith and Szathmary (1995) and compares their probabilities with their estimated times of occurrence.   He also discusses geochemical transitions relevant to biology, although they do not form part of the strictly biological evolutionary sequence based upon information.

**Critical Step Model for Evolution**

The critical step model assumes that there is an ordered sequence of transitions which govern the pace of evolution. Any intermediate steps or stages are assumed to occur rapidly without increasing the time between steps. The critical steps are assumed to occur stochastically with uniform, small, but unequal probabilities, $\lambda_1 \ldots \lambda_n$ where $\lambda_n << 1 / t_h$ where $t_h$ is the habitable lifetime of the planet. One alternative is a "long, slow fuse" model, which allows for a much larger sequence of more likely events.

The probability density for the *m*-th critical step in an *n* critical step model is:

$$P_{m/n}(t) = \frac{n!}{(n-m)!(m-1)!} \frac{t^{m-1}(t_h - t)^{n-m}}{t_h}$$

(1)

The expectation time for the *m*-th critical step in an *n* critical step model is:

$$<t_{m/n}> = \frac{m}{n+1} t_h$$

(2)

i.e. the spacing of the steps will therefore tend to be even throughout the habitability period, although they will be subject to statistical variation.

Watson adopted 4 Ga ago as the start of habitability, corresponding roughly to the period of late, heavy bombardment. The end of habitability is taken as 1 Ga in the future, corresponding to the time when solar luminosity and atmospheric models predict an extreme temperature rise resulting from faster weathering of rocks to carbonates and subsequent $CO_2$ loss. The probability density functions for the n-step models with n = 1,2 .. 5 are the polynomials shown in Table 2.

The fit between the probability density distributions in the 7-step model and the estimated times at which the critical steps actually occurred is poor. The main cause of the discrepancy is that prokaryotic life emerged relatively soon after habitability. The first three steps are much more closely spaced than the subsequent steps, in disagreement with the expectation of closer spacing predicted by the model. Watson shows that the fit can be significantly improved by assuming that the emergence of prokaryotic life only required a single initial step in a 5-step model.

| $P_{m/n}(t)$ | m = 1 | m = 2 | m = 3 | m = 4 | m = 5 |
|---|---|---|---|---|---|
| n = 1 | $\frac{1}{5}$ | | | | |
| n = 2 | $\frac{2}{5} - \frac{2}{25} t$ | $\frac{2}{25} t$ | | | |
| n = 3 | $\frac{3}{125} (5-t)^2$ | $\frac{6}{125} t (5-t)$ | $\frac{3}{125} t^2$ | | |
| n = 4 | $\frac{4}{625} (5-t)^3$ | $\frac{12}{625} t (5-t)^2$ | $\frac{12}{625} t^2 (5-t)$ | $\frac{4}{625} t^3$ | |
| n = 5 | $\frac{1}{625} (5-t)^4$ | $\frac{4}{625} t (5-t)^3$ | $\frac{6}{625} t^2 (5-t)^2$ | $\frac{4}{625} t^3 (5-t)$ | $\frac{1}{625} t^4$ |

Table 2  Probability Density Functions for 1 to 5 Step Models
(habitability lifetime $t_h$ = 5 Ga)

| $<t_{m/n}>$ | m = 1 | m = 2 | m = 3 | m = 4 | m = 5 |
|---|---|---|---|---|---|
| n = 1 | 5/2 | | | | |
| n = 2 | 5/3 | 10/3 | | | |
| n = 3 | 5/4 | 10/4 | 15/4 | | |
| n = 4 | 1 | 2 | 3 | **4** | |
| n = 5 | 5/6 | 10/6 | 15/6 | 20/6 | 25/6 |

Table 3 Expectation Time for Critical Transitions, $<t_{m/n}>$ in Ga
(habitable lifetime $t_h$ = 5 Ga)

Considering the final transition alone, the emergence of humans after 4 Ga of a 5 Ga habitable lifetime on Earth implies that the $n^{th}$ out of $n$ transitions occurred after 4 Ga. Table 3 shows the expectation times derived from equation (2) and indicates that this is most likely for n = 4, since $m = n = 4$ gives $<t_{4/4}>$ = 4 Ga. An alternative to a single step leading to prokaryotes is the suggestion that the timescale for evolution might be extended back by starting evolution earlier (Carter 2008, Watson 2008). Evidence of water on Earth 4.3 Ga ago (Valley at al. 2002) could signify the start of habitability. Meteorites from Mars suggest that life might have originated there and migrated to Earth (Treeman et al., 2000). If the origin of life occurred elsewhere in the Solar System, it must have been within the past 4.6 Ga. This extension of 0.6 Ga to the time over which life evolved helps to resolve the discrepancy between the model and the estimated transition times, but it does not alter the conclusion that the optimum number of critical transitions is less than seven. Carter (2008) suggests an optimum of six transitions if early evolution began on Mars, but stops short of proposing any earlier steps towards life.

**Origins within our Galaxy**

Our Galaxy appears to be almost as old as the Universe itself, since the oldest known halo star is estimated at 13.2 Ga old (Frebel, 2007). This ancient star is metal-poor and lacking in the heavy elements needed for planetary formation, while the oldest brown dwarfs, intermediates between stars and planets, are estimated to be 10 Ga old (Schilbach and Röser, 2009). The Galactic thin disc, where conditions are more favourable for life, is estimated to be between 6.5 and 11.1 Ga old (Del Peloso 2005).

The concept of a Galactic Habitable Zone, a region within the Galaxy favourable for the development of life, was first introduced by Gonzalez (2001) and expanded upon in Gonzalez (2005). Lineweaver et al. (2004) identified the present GHZ as an annular region between 7 and 9 kpc, which has moved outwards from the Galactic centre and is composed of stars that formed between 8 Ga and 4 Ga ago. His model of Galactic evolution and habitability considered the need for adequate star formation, heavy elements for planets to form and time for life to evolve, together with an absence of supernovae. Prantzos (2008) concludes that that habitability is far less clear-cut, because the factors involved are poorly understood. In particular, the effect of heavy metals on planet formation, hot Jupiters on terrestrial planet survival and supernovae radiation on life are extremely uncertain. His own model predicts a narrow annulus for Galactic habitability starting at 3 kpc around 13 Ga ago which then expands as it moves outwards, until it eventually encompasses the whole Galaxy within the lifetime of the Solar System and has a peak probability of Earth-like planets with life approaching 0.1 at 10kpc from the Galactic centre today. More recent work (Mattson, 2009) has modeled the build-up of carbon in the Galaxy predicting a peak at ~5.5 kpc around 8 Ga ago and at the solar distance ~9 kpc around 5 Ga ago, close to the time when the Earth formed.

Some of the necessary ingredients and conditions for life may therefore have existed within the Galaxy, albeit at different locations, for many billions of years before the formation of the Earth. Based upon current understanding, it is reasonable to consider the possibility of life originating elsewhere in the Galaxy before it began on Earth. Proponents of the Rare Earth Hypothesis (Brownlee and Ward, 2000) may place many further conditions upon the emergence of complex life, but we are concerned with the earliest, simple stages.

Table 4 summarises some of the timings for key events and transitions through the history of the Universe and the Earth. Although many may be regarded as controversial, they are used as best estimates in subsequent calculations. The origin of life in the Galaxy, or more specifically the start of Galactic habitability, is taken to have occurred $8.5 \pm 5$ Ga ago, where the central value of 8.5 Ga reflects the work of Lineweaver et al (2004), Prantzos (2008) and Mattson (2009) and the range of uncertainty allows for the most extreme conceivable values from the first stars in the Universe to the origin of life on Earth.

| Time in past (Ga) | Time since habitability on Earth (Ga) | Critical Event / Transition | Type |
|---|---|---|---|
| 13.7 | | Big Bang | C |
| 13.7 - | | nucleosynthesis of H/He after ~100s | C |
| 13.7 - | | atoms formed after 0.3 My (CMBR) | C |
| 13.6 | | first stars expected | C |
| 13.2 [1] | | first (halo) stars in Galaxy observed | A C |
| [13.5] **8.5** [3.5] [2] | | existence of Galactic Habitable Zone | A |
| [6.5] 8.3 [11.1] [3] | | Galactic thin disc formed | A |
| 4.57 | | formation of Sun | A E |
| 4.55 | | formation of Earth | E |
| [4.4] **4.3** [4] | | liquid water on Earth | E |
| [4.1] **4.0** [3.8] [5] | 0.0 | late heavy bombardment (Earth habitability) | L |
| | 0.2 | proto-cells (molecules in compartments) | B (1) |
| | 0.4 | chromosomes (RNA) | B (2) |
| [3.85] **3.5** - 3.0 [2.6] | 0.5 | emergence of prokaryotes (life) | B (3) |
| [3.2] 2.8 – 2.6 | | oxygen photosynthesis | E |
| 2.2 – 2.0 | | atmosphere becomes oxidising | E |
| [2.7] **1.9** – 1.2 [0.8] | 2.1 | prokaryotes to eukaryotes | B (4) |
| | 2.5 | asexual to sexual populations | B (5) |
| [1.9] **1.2** – 0.6 | 2.8 | protists to animals/plants/fungi (cell differentiation) | B (6) |
| 0.55 | | microscopic to macroscopic life (Cambrian) | E |
| 0.0001 – 0.0002 | 4.0 | primates to man | B (7) |
| (1.0) in future | 5.0 | predicted $CO_2$ loss (inhabitability) | B |

C = Cosmology     A = Astronomy / Galaxy     E = Earth Science / Geology
L = Lunar Science     B = Biological Information
[extreme estimates]     **best estimates**

[1] Frebel (2007)   [2] See preceding text   [3] Del Pelosa (2005)   [4] Valley et al (2002)
[5] Cohen et al. (2000)

(1) to (7) Critical Evolutionary Transitions
(Maynard-Smith and Szathmary 1995, Watson 2008)

Table 4  Astrobiological Timings

**Extending the Timescale in the Critical Transition Model**

Calculations of the critical transition model have been carried out using the Maple computer algebra system. Maple was used to plot and integrate the probability density distributions defined by equation (1) and hence determine the probability that a transition would occur before its estimated time. Since the distributions are polynomial functions, no numerical techniques have been required to perform the integrations. Figure 1 shows the probability density distributions for the 7-step and 5-step models with the best estimates for the actual transition times marked on each curve by a cross. For the intermediate curves a cross towards the peak of the distribution suggests a good fit. The integrated area below the probability density curve to the left of the cross provides a more precise measure for establishing the goodness of fit.

The adoption of the values in Table 4, which differ within reasonable limits from those adopted by Watson (2008), yields different probabilities that a step occurred at or before the observed time, but confirms his conclusions. A 7-step model fits poorly (probabilities 25%, 10%, 3%, 33%, 23%, 11%, 21% respectively ) while a 5-step model fits much better (probabilities 41%, 70%, 50%, 27%, 33% respectively), where values closer to 50% imply a more acceptable model. The essential distinguishing factor between the models is the rapid emergence of life, which is allowed for in the 5-step model by combining the first three transitions of the 7-step model into one.

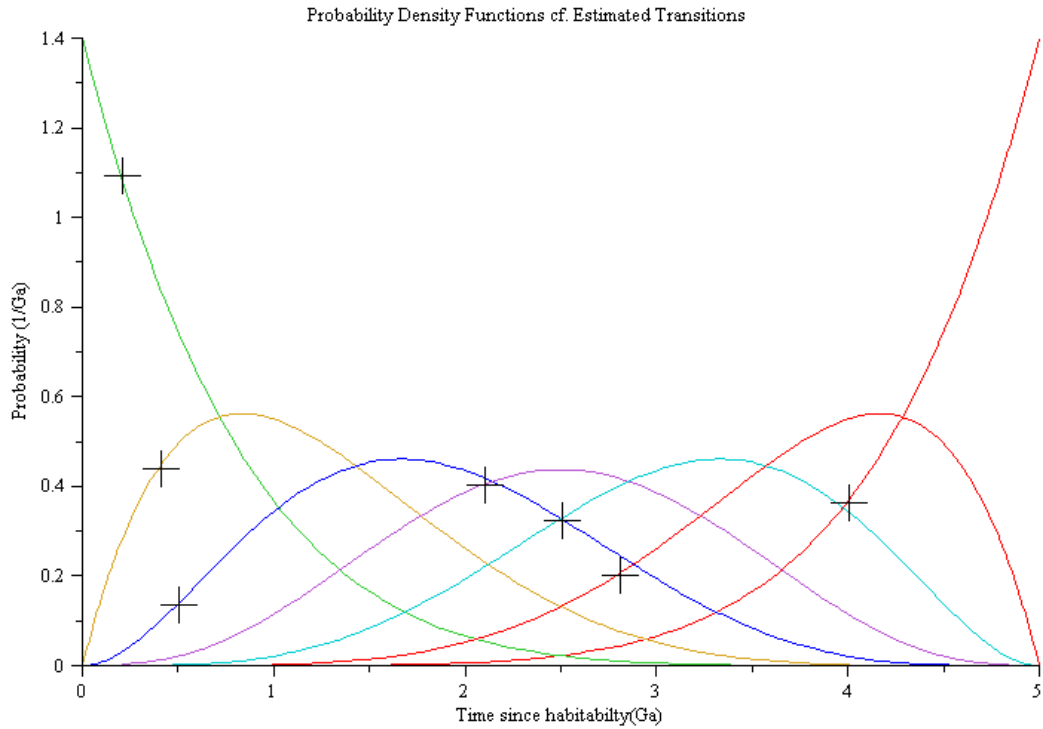
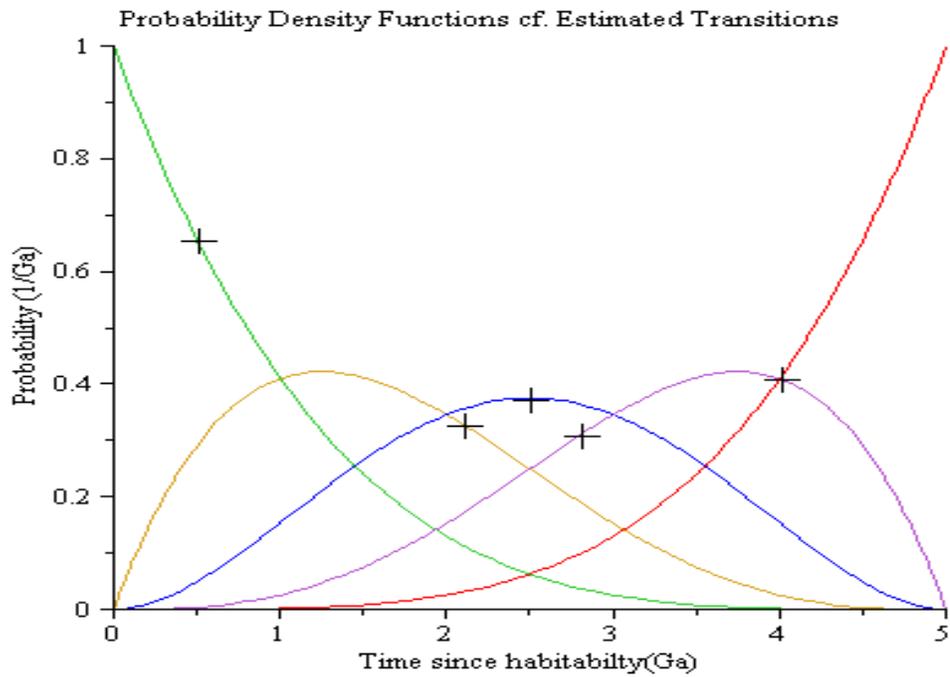

Figure 1  Critical Transition Models for (a) 7-steps (b) 5-steps
Compared with Estimated Transition Times (start of habitability = 4 Ga ago)

An alternative explanation is that the earlier transitions have occurred elsewhere in the Solar System or beyond. By considering the goodness of fit

$$F = \frac{\sum_{i=1}^{n} \sqrt{(P_i - 50)^2}}{n} \quad (3)$$

where

$P_i$ = probability of the $i^{th}$ step occurring before its actual estimated time (%)
$n$ = total number of steps

we have examined how well the model performs when different numbers of steps and earlier transitions on extraterrestrial locations are allowed. If all the probabilities were 50% , giving $F = 0$, the fit would be regarded as ideal. Figure 2 shows that $F$ is minimized for n = 5 and that the 5-step model does indeed fit the data best for purely Earth-based evolution.

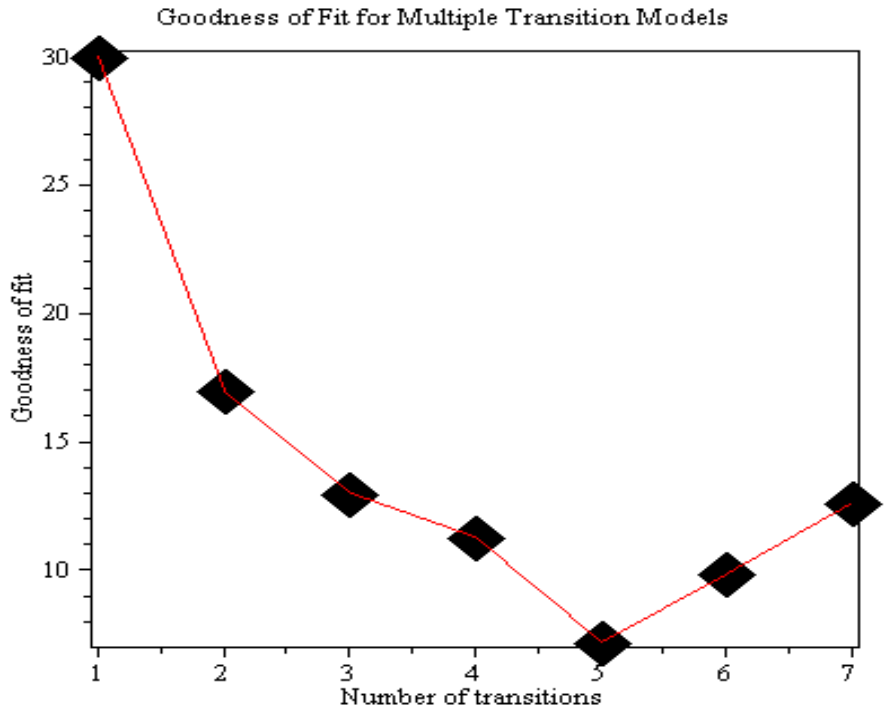

Figure 2  Optimisation of the Number of Transitions in the Critical Step Model

If these earlier transitions have taken place "off-site" they can be re-incorporated into the model by extending the timescale. Given that this needs to be longer than the age of the Solar System to make any significant difference to the model, a time of 8 Ga ago, corresponding to the estimated age of the thin disc and start of Galactic habitability according to the Lineweaver (2004) model, is adopted. The probability density distributions and estimated transition times are shown in Figure 3, where the probabilities for the 5 terrestrial transitions become 77%, 84%, 70%, 46%, 44%. The first two extraterrestrial transitions need to be excluded from the goodness of fit, since there is no sensible estimate for their times.

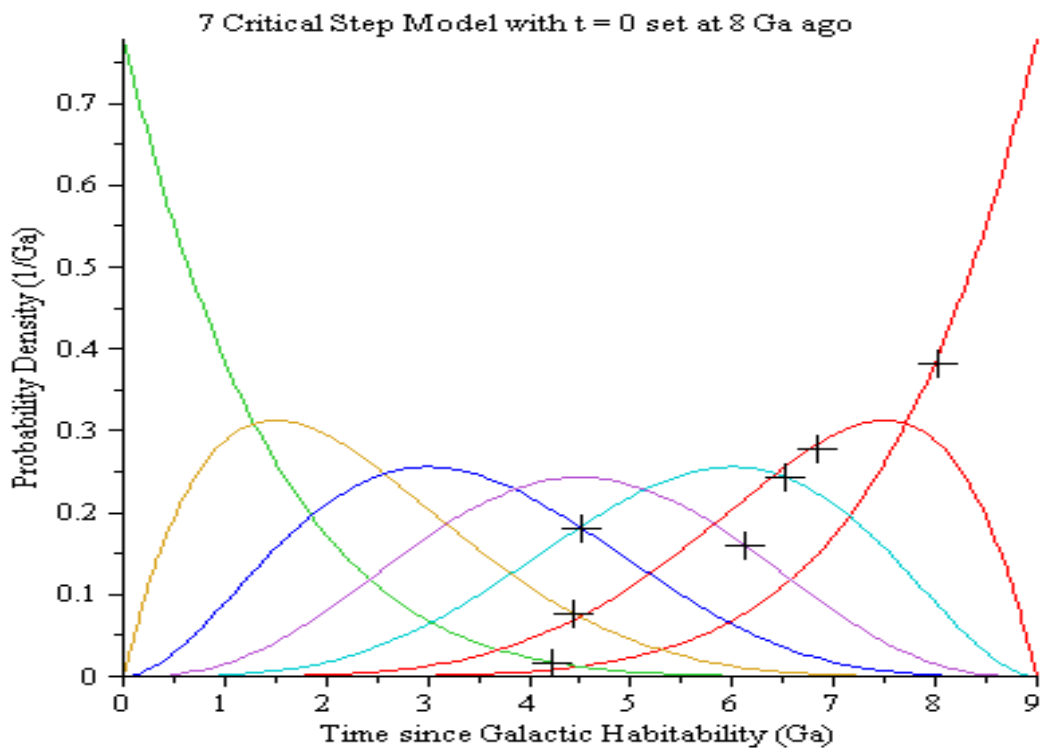

Figure 3 Critical Transition Models for 7-steps Compared with Estimated Transition Times (start of habitability = 8 Ga ago)

During the habitability period the first two transitions would have taken place on an extraterrestrial location and the final five on Earth, where we assume that the migration to Earth is not a critical step. However we can consider the consequences if the physical transfer to Earth is regarded as another critical step, resulting from an unlikely event over a long timescale? For example, a randomly occurring collision of a suitable comet or meteorite with Earth might be necessary. Alternatively, one collision or event might be required to initiate a transfer from one location and another to complete it by collision with the Earth. Lithopanspermia, an idea first suggested by Lord Kelvin in 1871, has been well discussed in subsequent years.  Some calculations have suggested that interstellar panspermia is unlikely (Melosh 2003), but viable mechanisms have subsequently been suggested (Napier 2004, Wallis and Wickramasinghe 2004).  The effect of adding one or two critical transitions to the model is shown in Figure 4.

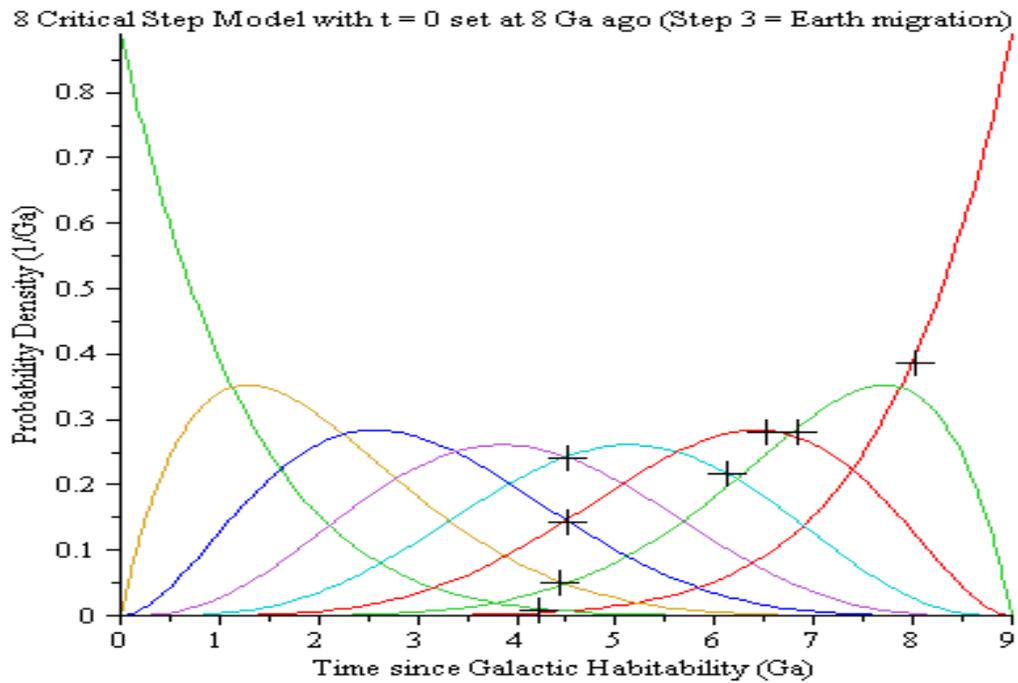
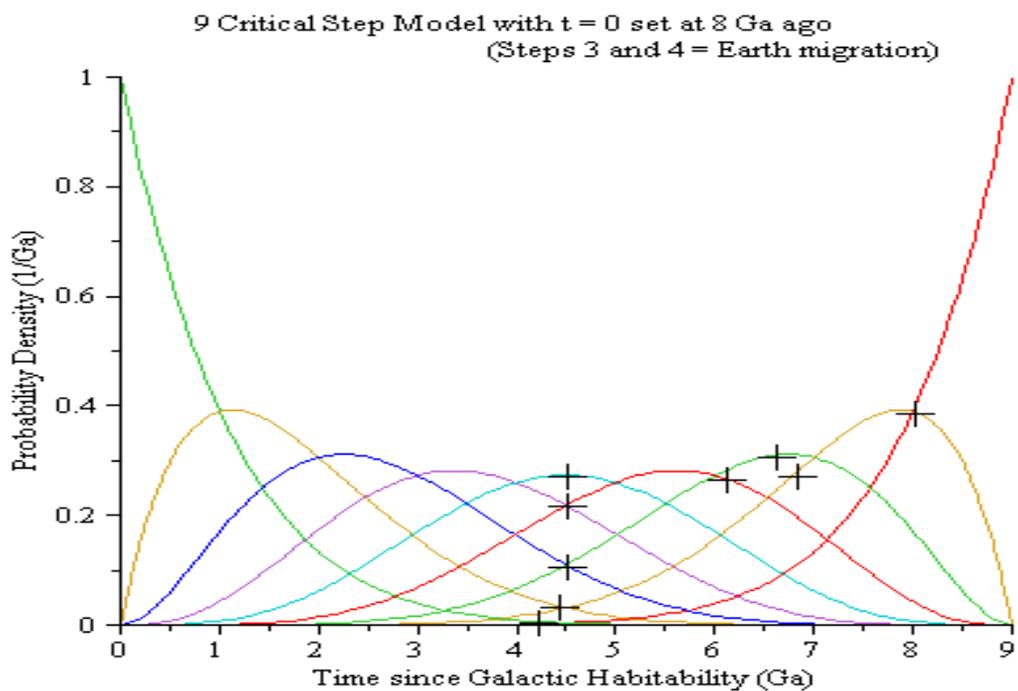
Figure 4 Critical Transition Models for (a) 8-steps (b) 9-steps
Compared with Estimated Transition Times (start of habitability = 8 Ga ago)

The best fit is obtained with the 9-step model, i.e. by having two Earth migration transitions (Figure 5a). By reducing the start of habitability to 6 Ga ago, a similar goodness of fit can be achieved without any Earth migration transitions (Figure 5b). Any further reduction in the start of habitability towards the age of the Solar System (Mars) decreases the goodness of fit (Figure 5c).

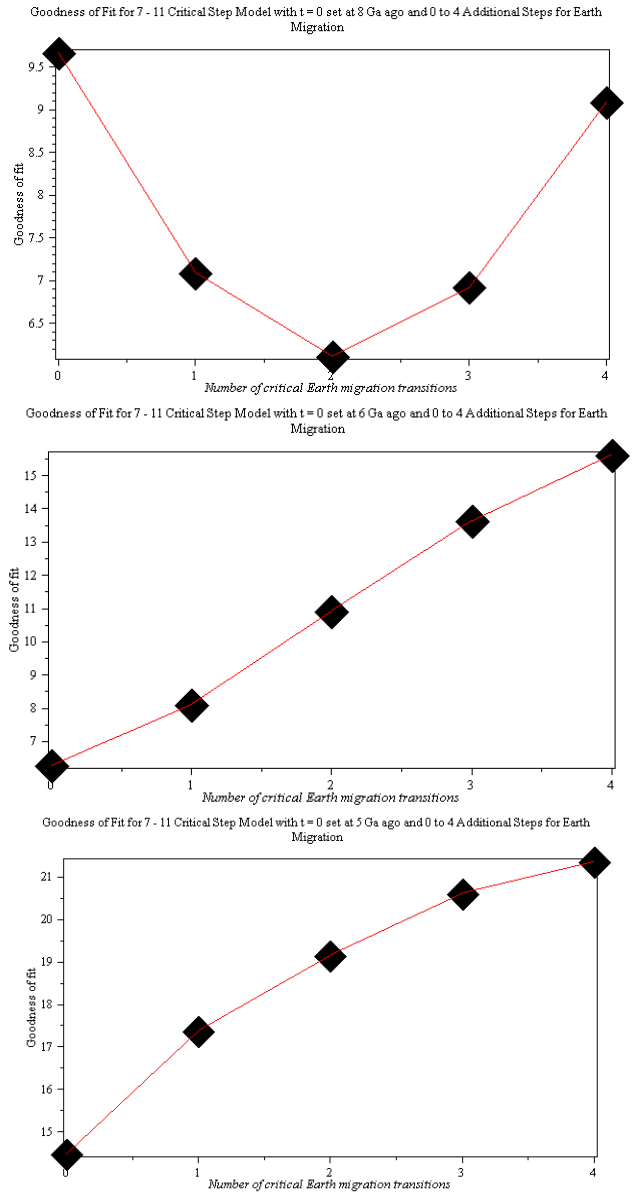

Figure 5 Model Dependence upon the Number of Earth Migration Transitions for Start of Habitability at (a) 8 Ga (b) 6 Ga (c) 5 Ga

**Discussion**

The equal spacing property of the critical transition model can be used to explain what is happening. If the transitions 1 - 7 are equally spaced upon the x-axis together with 0 = start of Earth habitability and 8 = end of Earth habitability and the estimated times from Table 4 are plotted, then a best fit line through the origin can be used for comparison (Figure 6). The fit is poor because of the rapid early transitions.

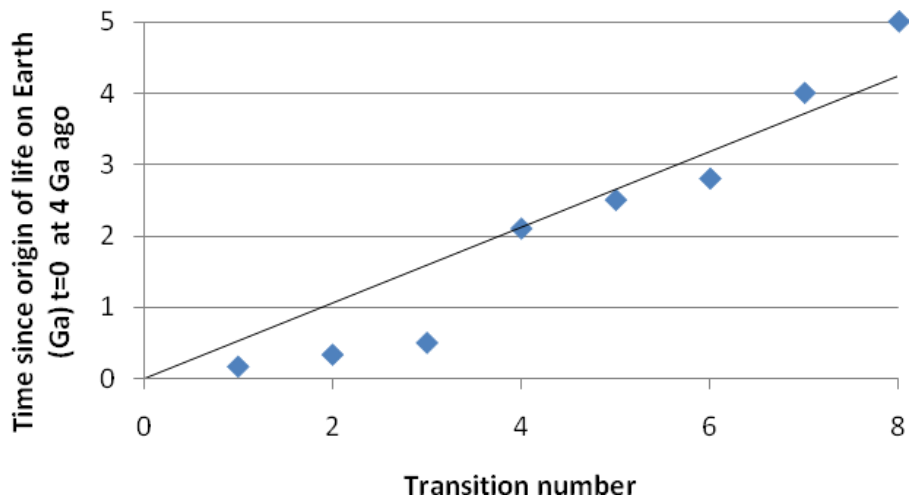

Figure 6 The Enforcement of Critical Spacing in a 7-Step Evolutionary Model

If the first two transitions are removed and the best fit line is no longer required to pass through the origin, the y-intercept of -1.6 Ga gives an estimate for the origin of life, i.e. around 5.6 Ga ago (Figure 7). This corresponds to the value of 6 Ga identified for the start of habitability when there were no Earth migration transitions. If additional transitions are introduced for Earth migration, then the y-intercept, will decrease and life emerge further back in the past.

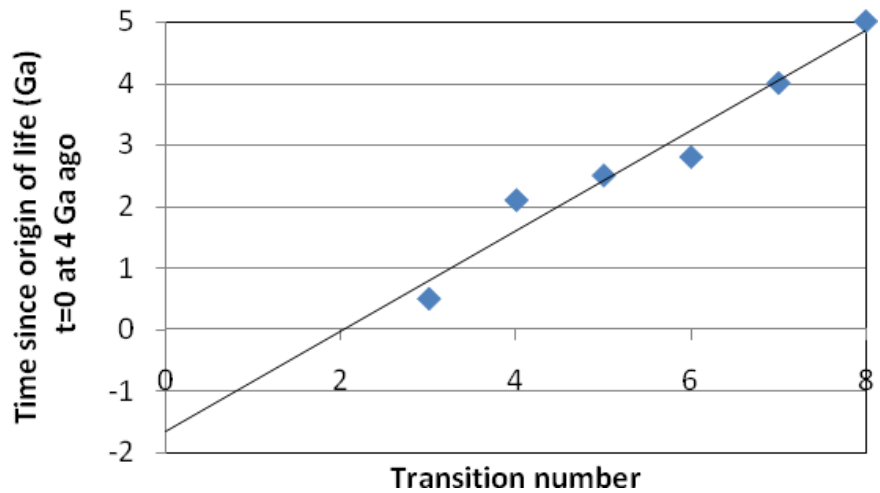

Figure 7 The Enforcement of Critical Spacing in a 5-Step Evolutionary Model

Lépine et al (2003) have shown that resonance with the spiral gravitation field, can cause stars to move radially through the Galaxy over distances of 2-3 kpc in significantly less than 1 Ga. Roškar et al. (2008) have modeled this stellar migration and considered its astrophysical implications. A serious possibility is that it could account for the migration of simple life from closer to the Galactic centre. Haywood (2009) have analysed solar neighbourhood stars showing that around 50% have come from elsewhere, primarily by migration from the inner disc, where Galactic habitability was greater at an earlier time. Not only does the Galactic Habitable Zone move radially outwards from the Galactic centre according to most models (Lineweaver, 2004, Prantzos 2008), but so do a significant number of stars. Thus from both a dynamic and chemical perspective, it may be possible for simple life to migrate from the inner disc of the Galaxy on realistic timescales and hence be included within the critical transition model. Wesson (2010) argues that, despite viable mechanisms for distributing organic material throughout the Galaxy, life could not survive the UV and cosmic ray bombardment. He does though suggest that necropanspermia, the transfer of information via damaged biological molecules, is feasible.

**Conclusions**

A simple stochastic model for evolution, based upon the need to pass a sequence of $n$ critical steps (Carter 1983, Watson 2008) has been applied to both terrestrial and extraterrestrial origins of life. In the former case, the time at which humans have emerged during the habitable period of the Earth suggests a value of $n = 4$.

Progressively adding earlier evolutionary transitions (Maynard Smith and Szathmary, 1995) gives an optimum fit when $n = 5$, implying either that their initial transitions are not critical or that habitability began around 6 Ga ago. The origin of life on Mars or elsewhere within the Solar System is excluded by the latter case. Carter (2008) in reviewing his original work, argues for a 5-step model for terrestrial evolution and a 6-step model if Mars evolution is included. If extra transitions are included to allow for Earth migration by simple life, the start of habitability needs to be at an even earlier time. Our present understanding of Galactic habitability and dynamics does not exclude this possibility. Each additional migration transition pushes back the start of habitability by around 1 Ga.

The Drake equation has long been used as a touchstone for astrobiology. Its mix of astrophysical, biological and social science, with an increasing level of uncertainty, continues to provoke discussion (Burchell, 2006). The critical transition model addresses an issue for any Drake-type equation, e.g. the Rare Earth equation (Brownlee and Ward, 2000). If any one of the factors is zero, then the result is zero. The critical transition model breaks down the crucial biological steps, typically only represented by one or two factors, and introduces time dependence. In particular, it provides a reasonable fit to the timescales for the emergence of life on Earth. Furthermore, if the possibility of life originating beyond the Earth is considered, it can provide useful clues as to the time that life might have emerged.

Whether the assumptions of the model and the identification of the transitions are reasonable is open to debate. For example, the likelihood of critical transitions could be governed by causal links between astrophysical and biological events. Transitions might occur relatively quickly with a high probability if life emerges through stages of Galactic punctuated equilibrium separated by irregular catastrophic (catalytic?) events in the Galaxy such as gamma ray bursts or supernovae as suggested by Circovic (2009). The Earth is certainly not a closed system, being exposed to electromagnetic radiation, neutrinos and cosmic rays from beyond the Solar System and material from within it. The extent to which the Earth is an open system and the degree of correlation between astrophysics and biology are key issues, which "grand" astrobiological models can help to resolve. Critical transition models, incorporating both biological and astrophysical concepts, provide a standard against which other models can be judged.

# References

Barrow, J.D. and Tipler, F.J., (1986). *The Anthropic Cosmological Principle*. Oxford University Press.

Burchell, M.J., (2006). *W(h)ither the Drake Equation?*. International Journal of Astrobiology, 5, 243-250

Carter, B., (1983). *The Anthropic Principle and its Implications for Biological Evolution.* Phil. Trans. R. Soc. Series A, v. 310, p. 347-363

Carter, B., (2008). *Five- or Six-Step Scenario for Evolution?.* International Journal of Astrobiology, 7:177-182, Cambridge University Press

Cirkovic, M.M., Vukotic, B., Dragicevic, I., (2009). *Galactic Punctuated Equilibrium: How to Undermine Carter's Anthropic Argument in Astrobiology.* Astrobiology, vol 9, no. 5, 491 - 581

Cohen, B.A. et al., (2000). *Support for the Lunar Cataclysm Hypothesis from Lunar Meteorite Impact Melt Ages*. Science 290, (5497):1754-1755

Del Peloso, E. F., (2005). *The Age of the Galactic Thin Disk from Th/Eu Nucleocosmochronology.* A&A, 440: 1153

Flambaum, V.V., (2003). Comment on *Does the Rapid Appearance of Life on Earth Suggest that Life is Common in the Universe?* Astrobiology 3, 237-239

Frebel, A., (2007). *Discovery of HE 1523-0901, a Strongly r-Process-enhanced Metal-poor Star with Detected Uranium*, Ap. J. 660: L117

Hanson, R., (1998). *Must Early Life Be Easy? The Rhythm of Major Evolutionary Transitions.* http://hanson.gmu.edu/hardstep.pdf

Haywood, M., (2009). *Radial Mixing and the Transition between the Thick and Thin Galactic Discs.* MNRAS, Volume 388 Issue 3, 1175 - 1184

Gonzalez, G. et al., (2001). *The Galactic Habitable Zone: Galactic Chemical Evolution.* Icarus, 152, 185 - 200

Gonzalez, G. ,(2005). *Habitable Zones in the Universe.* Origins Life Evol. Biospheres **33** 555

Lépine, J. R. D., Acharova, I. A. and Mishurov, Yu. N. (2003). *Corotation, Stellar Wandering, and Fine Structure of the Galactic Abundance Pattern.* Ap. J., 589:210-216

Lineweaver, C. H., Fenner, Y. and Gibson, B. K. (2004). *The Galactic Habitable Zone and the Age Distribution of Complex Life in the Milky Way.* Science, 303, 59-62.

Martin, W, et al., (2008). *Hydrothermal Vents and the Origin of Life*. Nature Reviews Microbiology, 6, 805 - 814

Mattson, L., (2009). *On the Existence of a Galactic Habitable Zone and the Origin of Carbon.* Swedish Astrobiology Meeting, Lund http://videos.nordita.org/conference/SwAN2009/Mattsson.pdf

Maynard Smith, J. and Szathmary, E., (1995). *The Major Transitions in Evolution.* Oxford University Press, ISBN 019850294X

Melosh, H.J., (2003). *Exchange of Meteorites (and Life?) Between Stellar Systems*. Astrobiology, 3(1): 207-215

Napier, W.M. (2004). *A Mechanism for Interstellar Panspermia*. Monthly Notices of the Royal Astronomical Society, 348, 1, 46-51

Prantzos, N, (2008). *On the "Galactic Habitable Zone".* Space Sci Rev 135:313

Rivkin, A. S. & Emery, J. P. *Detection of ice and organics on an asteroidal surface* Nature 464, 1322-1323 (2010).

Roškar, R., Debattista, V.P., Quinn, T.R., Stinson, G.S. and Wadsley, J., (2008). *Riding the Spiral Waves: Implications of Stellar Migration for the Properties of Galactic Discs*. Ap. J, 684: L79–L82

Schilbach E., Roeser S., Scholz R.D., (2009). *Trigonometric Parallaxes of Ten Ultracool Subdwarfs.* A&A, 493, 27

Treeman, A.H. et al., (2000). *The SNC Meteorites are from Mars*. Planetary Space Science. 48 (12-14): 1213-1230

Valley, J.W. et al. (2002). *A Cool Early Earth*. Geology 30: 351-354

Wallis, M.K. and Wickramasinghe, C.R. (2004). *Interstellar Transfer of Planetary Microbiota*. Monthly Notices of the Royal Astronomical Society, 348, 1, 52-61

Ward, P. D., and Brownlee, D., (2000). *Rare Earth: Why Complex Life is Uncommon in the Universe*. Copernicus Books, New York (Springer Verlag). ISBN 0-387-98701-0.

Watson, A. J. ,(2008). *Implications of an Anthropic Model for the Evolution of Complex Life and Intelligence.* Astrobiology J. v. 8, p 1-11.

Wesson, P.S., (2010). *Panspermia, Past and Present: Astrophysical and Biophysical Conditions for the Dissemination of Life in Space*. Space Sci Rev, Springer ISSN 1572-9672 (online)